# Towards a Software Architecture Maturity Model for Improving Ultra-Large-Scale Systems Interoperability

S. Shervin Ostadzadeh,
*Computer Eng. Dept., Science and Research Branch,
Islamic Azad University, Tehran, Iran
Email: ostadzadeh@srbiau.ac.ir

Fereidoon Shams
Computer Eng. Dept., Shahid Beheshti University
Tehran, Iran
Email: f_shams@sbu.ac.ir

*Abstract*. **For the last two decades, software architecture has been adopted as one of the main viable solutions to address the ever-increasing demands in the design and development of software systems. Nevertheless, the rapidly growing utilization of communication networks and interconnections among software systems have introduced some critical challenges, which need to be handled in order to fully unleash the potential of these systems. In this respect, Ultra-Large-Scale (ULS) systems, generally considered as a system of systems, have gained considerable attention, since their scale is incomparable to the traditional systems. The scale of ULS systems makes drastic changes in various aspects of system development. As a result, it requires that we broaden our understanding of software architectures and the ways we structure them. In this paper, we investigate the lack of an architectural maturity model framework for ULS system interoperability, and propose an architectural maturity model framework to improve ULS system interoperability.**

## 1. Introduction

Software engineering faces many challenges at the present time. Nevertheless, fundamental disparities between the current understanding of software and software development at the scale of Ultra-Large-Scale (ULS) [29] software-intensive systems remains one important challenge, which introduces critical constraints for effective achievement of the software engineering goals in a technical and economical manner. This is due to the fact that proper development of ULS systems has substantial impact on software engineering activities.

As systems grow larger and more complex to become ULS systems, new requirements for software architectures emerge. The software architecture of a program or computing system is the structure(s) of the system, which comprise software elements, the externally visible properties of those elements, and the relationships among them [2]. Based on this definition, it is inferred that software architecture characterizes the structure of a system. In general, architecture is the fundamental organization of a system embodied in its components, their relationships to each other, and to the environment, and the principles guiding its design and evolution [17].

According to the ISO 15704 standard [16], an architecture represents a description of the basic arrangement and connectivity of parts of a system (either a physical or a conceptual object or entity), which is expected to create a comprehensive overview of the entire system when put together [8]. It should be noted that handling this large amount of information is quite challenging and needs a well-developed framework. The problem is even intensified in the case of ULS systems, due to their scale. So far, various Information Systems Architecture (ISA) frameworks have appeared in literature: Zachman framework [30,34], FEAF [9], TEAF [10], ToGAF [24], E2AF[28], and C4ISR [5,6] to name a few. Nevertheless, these frameworks fail to provide all the required support for ULS systems. Consequently, the inability of current ISA frameworks to meet these requirements necessitates a breakthrough research in the development of a ULS architectural framework [29].

In this paper, we present an architectural maturity model framework in ULS systems interoperability based on complex system theory. The proposed framework is assumed to be capable of addressing the requirements of such systems.

The rest of the paper is organized as follows. In Section 2, we present the required background and the problem definition. We introduce the ULS interoperability model based on complex system theory in Section 3. The ULS maturity models are discussed in Section 4. Finally, Section 5 summarizes the contributions and sets the direction for the future work.

## 2. Background

It has been observed that current approaches fail to fully define, develop, deploy, operate, acquire, and evolve ULS systems, as described in SEI report [29]. ULS systems are considered as cities or socio-technical ecosystems, while our current knowledge and practices are geared toward creating individual buildings or species. This inconsistency points out the research direction that is crucial for reaching a proper solution to develop ULS systems. The challenges that have to be addressed when developing a ULS system span three different areas: 1) Design and Evolution, 2) Orchestration and Control, and 3) Monitoring and Assessment [29].

### 2.1. Research context

The research work presented here addresses the design area related to "design and evolution". Fundamental to the design and evolution of a ULS system will be explicit attention to design across logical, spatial, physical, organizational, social, cognitive, economic, and other aspects of the system. Attention to design is also needed across various levels of abstraction involving hardware and software as well as procurers, acquirers, producers, integrators, trainers, and users. A key area of research in design is thus the need for design of all levels of a ULS system.

### 2.2. Why interoperability?

Broadly speaking, interoperability refers to coexistence, autonomy, and federated environment, whereas integration conventionally refers to the concept of coordination, coherence, and uniformization [8]. ULS systems go far beyond the size of current systems and system of systems by every measure, including, the number of the lines of code; the number of people using the system for different purposes; amount of data stored, accessed, manipulated, and refined; the number of connections and interdependencies among software components; and the number of hardware elements [29]. These are instances of 'Loosely coupled' systems. This means that the components in such systems can interact and are connected by a communication network; they can exchange services while continuing locally their own logic of operation. "Tightly-Coupled" indicates that the components are interdependent and cannot be separated. This is the case of a fully integrated system. Thus, two integrated systems are inevitably interoperable, however, two interoperable systems are not necessarily integrated.

### 2.3. Related work

Since the beginning of the last decade, the research work on architecture development is based on the improvements in enterprise interoperability frameworks. Generally, the main purpose of such frameworks is to provide an organizing mechanism so that concepts, problems, and knowledge on enterprise interoperability can be represented in a more structured way [8].

The LISI (Levels of Information Systems Interoperability) approach [6], developed by C4ISR Architecture Working Group (AWG) in 1997, is a framework to provide the US Department of Defense (DoD) with a maturity model and a process for determining joint interoperability needs, assessing the ability of the information systems to meet these needs, and selecting pragmatic solutions in addition to a transition path for achieving higher states of capability and interoperability.

The IDEAS interoperability framework [15] reflects the view that interoperability is achieved on multiple levels. These levels include inter-enterprise coordination, business process integration, semantic application integration, syntactical application integration, and physical integration.

The ATHENA Interoperability Framework (AIF) [1] is structured into three levels. The conceptual level is used for the identification of research requirements and the integration of research results. The applicative level is used for knowledge transfer regarding the application of integration technologies. The technical level is used for technology testing based on profiles and the integration of prototypes.

The E-health interoperability framework [22], which is developed by NEHTA (National E-Health Transition Authority) initiatives in Australia, brings together organizational, information, and technical aspects related to the delivery of interoperability across health organizations.

The European Interoperability Framework (EIF) [12,13] aims at supporting the European Union's strategy of providing user-centered eGovernment services. This is achieved by defining services as overarching set of policies, standards, and guidelines, which describe the way in which organizations have agreed, or should agree, to do business with each other.

In United Kingdom, the eGovernment Unit7 (eGU), has based its technical guidance on the eGovernment Interoperability Framework (e-GIF) [11]. e-GIF mandates sets of specifications and policies for any cross-agency collaboration as well as for e-government service delivery.

The NATO C3 Interoperability Environment (NIE) [20] encompasses the standards, products, and agreements adopted by the Alliance to ensure C3 interoperability. It serves as the basis for the development and the evolution of C3 Systems.

Layers of Coalition Interoperability (LCI) [31] is a framework for possible measures of merit to deal with the various layers of semantic interoperability in coalition operations.

System of Systems Interoperability (SOSI) [19] introduces three types of interoperability: 1) programmatic: interoperability between different program offices, 2) constructive: interoperability between the organizations that are responsible for the construction (and maintenance) of a system, and 3) operational: interoperability between the systems.

### 2.4. Research context

The scale of complexity and uncertainty in the design of ULS systems is so immense that resists the treatments offered by traditional interoperability methods. According to SEI report [29], ULS system complexity is a new perspective: "architecture is not purely a technical plan for producing a single system or closely related family of systems, but a structuring of the design spaces that a complex design process at an industrial scale will explore over time". Breaking up an architecture into design spaces and striving for a set of coherent and effective design rules would seem to imply a significant degree of control of the overall design and production process. Nevertheless, the design spaces, design rules, and organizations will be continually adjusting and adapting to both internal and external forces, which makes it difficult to handle them all.

The criticality of the research is justified by the fact that handling the large volume of information available in ULS systems is only feasible by utilizing a well-developed interoperability framework. A newly proposed framework is expected to broaden a traditional interoperability framework to include people and organizations; social, cognitive, and economic considerations; and design structures such as design rules and government policies.

This research work centers around the development of an architectural framework to improve the interoperability of ULS systems. We pose the question that given the issues with the design of all levels of ULS architectures, how can one organize and classify the types of information that must be created and used in order to improve the ULS interoperability?

### 3. Complex system theory

A complex system is a system composed of interconnected parts that, as a whole, exhibit one or more properties (behavior among the possible properties) not obvious from the properties of the individual parts [18]. The complexity of a system may be of one of the two forms: disorganized complexity and organized complexity [33].

The scale of ULS systems reveals some characteristics that are not seemingly visible in traditional systems [14,29]: (1) decentralization; (2) inherently conflicting, unknowable, and diverse requirements; (3) continuous evolution and deployment; (4) heterogeneous, inconsistent, and changing elements; (5) erosion of the people/system boundary; (6) normal failures; (7) new paradigms for acquisition and policy. These characteristics undermine current, widely used, information systems framework and establish the basis for the technical challenges associated with ULS systems.

**Table 1.** Complex systems and ULS systems similarities

| Complex Systems | ULS Systems |
|---|---|
| Difficult to determine boundaries | Erosion of the people/sys. boundary |
| May be open Low | Erosion of the people/sys. boundary |
| May have a memory | Continuous evolution & deployment |
| Dynamic network of multiplicity | Decentralization |
| May produce emergent phenomena | Inherently conflicting req. |
| Relationships are non-linear | Heterogeneous and inconsistent |
| Relationships contain fb. loops | Continuous evolution & deployment |

ULS systems are examples of disorganized complexity because disorganized complexity is a matter of a very large number of parts. Table 1 lists the similarities between the features of complex systems and their corresponding parts in ULS systems.

### 3.1. ULS interoperability model

As introduced in Section II.C, the SOSI [19] can be considered as a significant initiative for ULS systems interoperability. However, as mentioned in SEI report [29], people will not just be users of a ULS system, rather, they will be part of its overall behavior. In addition, the boundary between the system and user/developer roles will blur. Just as people who maintain and modify a city, may also reside in the city, in a ULS system, a person may act in the role of a traditional user, or in a supporting role as a maintainer of the system health, or as a change agent adding and repairing the functions of the system.

Assuming that people are part of a ULS system signifies that a new perspective has to be taken into account: culture. Figure 1 depicts an extension to the SOSI model in order to achieve ULS system socio-technical characteristics. The four layers of ULS interoperability model corresponds to the four layers of complex system theory model. In complex system theory, we can divide a system into four layers: 1) vital, 2) psyche, 3) social, and 4) cultural [32].

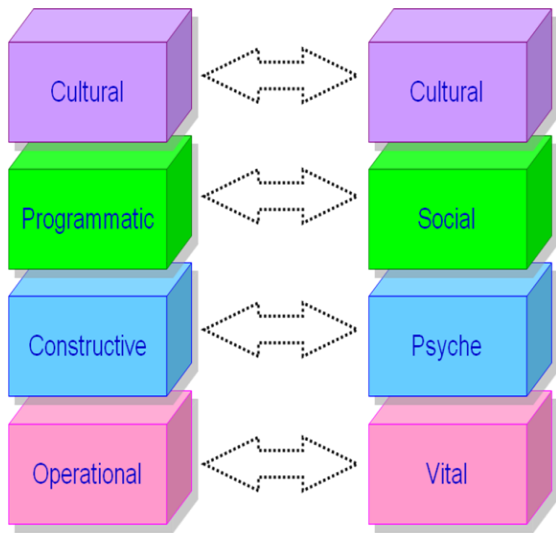

**Figure 1.** Alignment between ULS interoperability model and complex system theory [27]

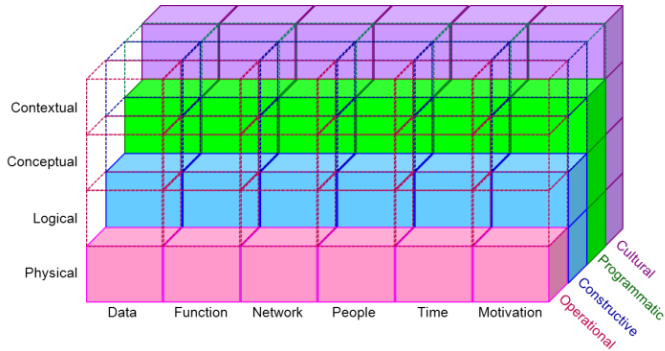

**Figure 2.** ULS interoperability framework (Blank cells are not supposed to be modeled.)

### 3.2. ULS interoperability framework

Zachman Framework (ZF) [34], originally proposed by John Zachman, is often referenced as a standard approach for expressing the basic elements of information system architecture, and is widely accepted as the main framework in ISA. Although some of today's successful ISA frameworks (including ZF) are used for enterprise systems architecture, the problem discussed in the previous section is inherently broader and deeper than current capabilities of ISA frameworks [3,4,7,21,23,25,26]. Figure 2 depicts our initiative proposed framework to improve interoperability based on complex system theory. In this work, we apply ZF as an initial start and try to enrich it by ULS Interoperability model to support the special characteristics of ULS interoperability. The proposed framework should be a spectrum of technologies and methods with software engineering, economics, human factors, cognitive psychology, sociology, systems engineering, and business policy.

### 4. Interoperability maturity model

Following the discussion in Section 3 and based on a systemic view of ULS interoperability framework, we identify five maturity levels of interoperability, as listed in Table 2. The transition from one level to a higher one entails the removal of interoperability barriers and the satisfaction of interoperability requirements. It is important to note that a lower interoperability maturity does not systematically mean a malfunction of the system. The maturity is only evaluated from the interoperability point of view and is not applicable for other purposes.

### 4.1. Level 0 (Isolated)

The initial maturity level of interoperability is characterized by isolated systems. In such systems, resources are not intended to be shared with others. System modeling and description are incomplete or even nonexistent. Generally, no interoperation is possible or desired. Communication remains mainly as manual exchange of information. Systems run standalone and they are not prepared for interoperability.

### 4.2. Level 1 (Operated)

At this maturity level, systems may fully integrate (note that this is in contrast to interoperate). All interactions happen in the operational layer, however interoperability remains very limited. Basic IT devices are connectable and electronic data exchange becomes feasible. Systems are generally defined and modeled separately.

**Table 2.** ULS interoperability maturity model

| Level | Name | Description |
|---|---|---|
| 0 | Isolated | No Interoperability. Systems work without any interaction. |
| 1 | Operated | Common operational layer. Systems share common data (M0). |
| 2 | Constructed | Common constructive layer. Systems share common model (M1). |
| 3 | Programmed | Common programmatic layer. Systems share common meta-model (M2). |
| 4 | Allied | Common cultural layer. Systems share common meta-meta model (M3). |

### 4.3. Level 2 (Constructed)

This level of maturity requires common models that enable a system to create and to make changes in its data so that to adhere to common formats. In addition, relevant standards are used as much as possible. Models remain platform-dependent. Nevertheless, models are used not only for modeling at design time, but also for execution at run time.

### 4.4. Level 3 (Programmed)

At this maturity level, systems are well organized to handle interoperability challenges. Interoperability capability is extended to heterogeneous systems, often in a networked domain. Although systems remain heterogeneous, meta-modeling is performed and mapping is generalized using meta-models. Systems are capable of interoperating with multiple heterogeneous partners.

### 4.5. Level 4 (Allied)

This level corresponds to the highest maturity level of interoperability. Systems are able to dynamically adjust themselves and modifications are carried out on the fly. Shared domain ontologies/strategies are generally existent. At this level, systems are able to interoperate with multi-lingual and multi-cultural heterogeneous partners. Additionally, all information becomes a subject of meta-meta model and can be adapted at runtime.

## 5. Conclusion

Achieving ULS interoperability involves changes to the way we define life, including acquisition practices and guidance, technologies, engineering and management practices, operational doctrines for both the usage and those who support the systems. Realizing this vision requires that we begin to define approaches and models in more concrete terms.

In this paper, an architectural maturity model based on complex system theory is proposed to improve ULS system interoperability. This allows software architects to model various aspects of ULS systems interoperability. The proposed model presents a classification schema for descriptive representation of a ULS system. The goal is that the framework be used to complement a full-structural schema within the ULS interoperability maturity model. In particular, this approach will enable architects to:

- classify the ULS maturity model interoperability;
- represent and analyze ULS levels of interoperability;
- work with others toward a complete and consistent set of interoperability models

As the future work, one is expected to propose an appropriate methodology to help increasing architectural maturity level in ULS systems.

## References

[1] ATHENA, Advanced Technologies for Interoperability of Heterogeneous Enterprise Networks and their Applications, FP6-2002-IST1, Integrated Project, 2003.

[2] L. Bass, P. Clements, R. Kazman, Software Architecture in Practice, 2nd Edition, SEI Series in Software Architecture, Addison-Wesley Professional, 2004.

[3] S. Blanchette, P. Clements, M. Gagliardi, J. Klein, U.S. Army Workshop on Exploring Enterprise, System of Systems, System, and Software Architectures (CMU/SEI-2008-TR-023), Software Engineering Institute (SEI), Carnegie Mellon University (CMU), Pittsburgh, PA, USA, 2009.

[4] P. Boxer, S. Garcia, "Enterprise Architecture for Complex System-of-Systems Contexts", 3rd Annual IEEE International Systems Conference, Vancouver, Canada, 2009.

[5] C4ISR Architecture Working Group (AWG), C4ISR Architecture Framework, Version 2.0, USA Department of Defense (DoD), 1997.

[6] C4ISR Architecture Working Group (AWG), Levels of Information Systems Interoperability (LISI), USA Department of Defense (DoD), 1998.

[7] P. Clements, "Exploring Enterprise, System of Systems, and System and Software Architectures", SEI Webinar, Carnegie Mellon University (CMU), Pittsburgh, PA, USA, 2009.

[8] D. Chen et al., "Architectures for enterprise integration and interoperability: Past, present and future", Comput Industry (Ind), 2008. doi:10.1016/j.compind.2007.12.016.

[9] Chief Information Officers (CIO) Council, Federal Enterprise Architecture Framework, Version 1.1, USA, 1999.

[10] Department of the Treasury, Treasury Enterprise Architecture Framework, Version 1, USA, 2000.

[11] eGU, eGovernment Unit, eGovernment Interoperability Framework (eGIF), version 6.1, UK, 2005.

[12] EIF, European Interoperability Framework, Brussels, 2004.

[13] EIF, European Interoperability Framework for PAN-European eGovernment Services, IDA Working Document, Version 4.2, 2004.

[14] G. Goth, "Ultra-Large System: Redefining Software Engineering", IEEE Software Journal, Vol. 25, Issue 3, pp. 91-94, 2008.

[15] IDEAS, Interoperability Development for Enterprise Application and Software Roadmaps, Annex 1—DoW, 2002.

[16] ISO, Industrial Automation Systems—Requirements for Enterprise-reference Architectures and Methodologies, ISO 15704, 2000.

[17] IEEE Standards board, Recommended Practice for Architectural Description of Software-Intensive Systems, IEEE-Std-1471-2000, 2000.

[18] C. Joslyn, L. Rocha, "Towards semiotic agent-based models of socio-technical organizations", Proceeding of AI, Simulation and Planning in High Autonomy Systems (AIS 2000) Conference, Tucson, Arizona, USA, pp. 70-79, 2000.

[19] E. Morris et al., System of Systems Interoperability, Technical Report, Carnegie Mellon University (CMU), Pittsburgh, PA, USA, 2004.

[20] NC3A. NATO C3 Technical Architecture Reference Model for Interoperability, NATO Consultation, Command, and Control Agency, 2003.

[21] NECSI, Characteristics of Systems of Systems, NECSI: Complex Physical, Biological and Social Systems Project, 2004.

[22] NEHTA, Towards an Interoperability Framework, Version 1.8, 2005.

[23] Office of the Deputy Under Secretary of Defense for Acquisition and Technology, Systems and Software Engineering: Systems Engineering Guide for Systems of Systems, Version 1.0, Washington, DC: ODUSD(A&T)SSE, USA, 2008.

[24] Open Group, The Open Group Architecture Framework (TOGAF), Version 9.0., USA, 2009.

[25] S. S. Ostadzadeh, F. Shams, S. A. Ostadzadeh, "A Method for Consistent Modeling of Zachman Framework. Advances and Innovations in Systems, Computing Sciences and Software Engineering", Springer, pp. 375-380, 2007. doi=10.1007/978-1-4020-6264-3_65

[26] S. S. Ostadzadeh, F. Shams, S. A. Ostadzadeh, "An MDA-Based Generic Framework to Address Various Aspects of Enterprise Architecture. Advances in Computer and Information Sciences and Engineering", Springer, pp. 455-460, 2008. doi:10.1007/978-1-4020-8741-7_81

[27] S. S. Ostadzadeh, B. Rezaei Rad, F. Shams, "An Interoperability Architectural Model based On Complex System Theory for the Ultra-Large-Scale Systems", Proceedings of Software Engineering and Application (SEA' 2011), International Association of Science and Technology for Development (IASTED), 2011.

[28] J. Schekkerman, Extended Enterprise Architecture Framework Essentials Guide, Version 1.5, Institute For Enterprise Architecture Developments (IFEAD), 2006.

[29] Software Engineering Institute (SEI), Ultra-Large-Scale Systems: Software Challenge of the Future, Technical Report, Carnegie Mellon University (CMU), Pittsburgh, PA, USA, 2006.

[30] John F. Sowa, John A. Zachman, "Extending and Formalizing the Framework for Information Systems Architecture", IBM Systems Journal, Vol. 31, No. 3, pp. 590-616, 1992.

[31] A. Tolk, "Beyond Technical Interoperability: Introducing a Reference Model for Measures of Merit for Coalition Interoperability", Proceedings of 8$^{th}$ ICCRTS, Washington, USA, 2003.

[32] S. Vakili, Complex Systems Theory, Shourafarin Publication, 2010.

[33] W. Weaver, "Science and Complexity", American Scientist 36: 536, 1948. (Retrieved on 2007–11–21.)

[34] John A. Zachman, "A Framework for Information Systems Architecture", IBM Systems Journal, Vol. 26, No. 3, pp. 276-292, 1987. (Reprinted in 1999: Vol. 38, No. 2-3, 1999.)

* Corresponding Author:
S. Shervin Ostadzadeh,
Faculty of Electrical and Computer Engineering,
Science and Research Branch, Islamic Azad University, Tehran, Iran,
Email: ostadzadeh@srbiau.ac.ir     Tel:+98-21-44869655